\documentclass[journal]{IEEEtran}
\IEEEoverridecommandlockouts
\usepackage{lineno}
\usepackage{hyperref}
\usepackage{cite}
\usepackage{amsmath,amssymb,amsfonts,cases} 
\usepackage{amsmath}
\usepackage{amsthm}
\DeclareMathOperator*{\argmax}{argmax}
\DeclareMathOperator*{\argmin}{argmin}
\usepackage{graphicx}
\usepackage{textcomp}
\usepackage{xcolor}
\usepackage{graphicx}
\usepackage{float}
\usepackage{subfigure}
\usepackage{amsmath,mleftright}
\usepackage{amsfonts,amssymb}
\usepackage{mathrsfs}
\usepackage{mathtools}
\usepackage{algorithm}
\usepackage{algorithmic}
\usepackage{bm}
\usepackage{multirow}
\usepackage{array}
\usepackage{amssymb}
\usepackage{amsmath}
\usepackage{cite}
\usepackage{url}
\usepackage{xcolor}
\usepackage{cite,graphicx,amsmath,amssymb}
\usepackage{subfigure}
\usepackage{fancyhdr}
\usepackage{mdwmath}
\usepackage{mdwtab}
\usepackage{caption}
\usepackage{amsthm}
\usepackage{setspace}
\usepackage{bm}
\usepackage{mathtools}
\usepackage{dsfont}
\usepackage{bbm}
\usepackage{framed}

\newtheorem{theorem}{Theorem}

\newtheorem{lemma}{Lemma}

\newtheorem{corollary}{Corollary}

\makeatletter
\newcommand{\biggg}{\bBigg@{3}}
\newcommand{\Biggg}{\bBigg@{3.5}}
\makeatother
\makeatletter
\allowdisplaybreaks[4]
\renewcommand{\maketag@@@}[1]{\hbox{\m@th\normalsize\normalfont#1}}%
\makeatother
\def\BibTeX{{\rm B\kern-.05em{\sc i\kern-.025em b}\kern-.08em
    T\kern-.1667em\lower.7ex\hbox{E}\kern-.125emX}}
    \expandafter\def\expandafter\normalsize\expandafter{%
    \normalsize%
    \setlength\abovedisplayskip{4pt}%
    \setlength\belowdisplayskip{4pt}%
    \setlength\abovedisplayshortskip{2pt}%
    \setlength\belowdisplayshortskip{2pt}%
}
\begin{document}
\title{Over-the-Air Computation via Segmented Waveguide-Enabled Pinching-Antenna Systems}
\author{Songnan~Gu, Hao~Jiang, Chongjun~Ouyang, Dian Fan,\\Yuanwei~Liu, \IEEEmembership{Fellow, IEEE}, and Arumugam~Nallanathan, \IEEEmembership{Fellow, IEEE}\vspace{-10pt}
\thanks{S. Gu, H. Jiang, C. Ouyang, and A. Nallanathan are with the School of Electronic Engineering and Computer Science, Queen Mary University of London, London, E1 4NS, U.K. (e-mail: s.gu@se23.qmul.ac.uk; \{hao.jiang, c.ouyang, a.nallanathan\}@qmul.ac.uk).} 
\thanks{D. Fan is with the China Academy of Information and Communications Technology, Beijing, 100191, China (email: fandian@caict.ac.cn).}
\thanks{Y. Liu is with the Department of Electrical and Electronic Engineering, The University of Hong Kong, Hong Kong (email: yuanwei@hku.hk).}}
\maketitle
\begin{abstract}
A segmented waveguide-enabled pinching-antenna system (SWAN)-assisted over-the-air computation (AirComp) framework is proposed. Three transmission architectures, namely segment selection (SS), phase-shifter-free segment aggregation (SA), and phase-shifter-enabled SA, are developed for uplink signal aggregation. For each architecture, low-complexity algorithms are developed to optimize the pinching-antenna placement and the per-segment phase shifts. Numerical results demonstrate the effectiveness of the proposed approaches and the superiority of SWAN over the conventional pinching-antenna system (PASS). It is shown that both SS and SA achieve lower computation mean-squared error than the conventional PASS, while segment-wise phase control further improves the performance of SA.
\end{abstract}

\section{Introduction}
Next-generation wireless networks are expected to support massive edge intelligence and latency-sensitive services \cite{Dang2020What6G}. These emerging applications require efficient multiple access mechanisms for large-scale distributed data aggregation \cite{Wang2024AirComp}. Over-the-air computation (AirComp) has therefore attracted significant attention since it exploits the waveform superposition property of wireless multiple access channels (MACs) to directly compute functions of distributed signals during transmission \cite{Zhu2021AirComp,Goldenbaum2013AnalogComp}. As a result, AirComp can significantly reduce communication latency and spectrum consumption for real-time intelligent applications such as federated learning and autonomous control \cite{wen2024survey}.

The performance of AirComp critically depends on accurate channel alignment among distributed users. Channel fading, blockage, and spatial channel variations can introduce severe aggregation distortion and degrade the computation accuracy \cite{Wang2024AirComp}. To improve channel alignment, recent studies have investigated several reconfigurable-antenna-enabled transmission architectures, including reconfigurable intelligent surfaces \cite{fang2021over,Zhang2022WorstCaseRISAirComp}, stacked intelligent metasurfaces \cite{stylianopoulos2026over}, and movable/fluid antennas \cite{Li2025AirComp2DMAA}. However, these approaches often rely on large passive surfaces or mechanical antenna movement, which may increase hardware complexity, signaling overhead, and energy consumption. Moreover, their spatial reconfigurability is typically limited to several wavelengths. Such a limited adjustment range is insufficient to effectively combat large-scale path loss and signal blockage.

As an alternative, the pinching-antenna system (PASS) was recently proposed for reconfigurable wireless communications \cite{suzuki2022pinching,Ding2025FlexibleAntenna}. PASS deploys pinching antennas (PAs) along dielectric waveguides to enable large-scale channel reconfiguration. Such an architecture can establish stable line-of-sight (LoS) links and bypass blockage regions. PASS is therefore regarded as a promising architecture for AirComp \cite{Lyu2025PASSAirComp}. Nevertheless, directly applying conventional PASS architectures to AirComp still faces a fundamental uplink challenge. Specifically, when multiple PAs are simultaneously activated on the same waveguide, inter-antenna radiation (IAR) arises since guided signals can be re-radiated by intermediate PAs before reaching the feed point \cite{Ouyang2026SWAN}. This effect introduces severe signal coupling among PAs. Moreover, a tractable yet physically consistent IAR signal model is still unavailable. Existing PASS-assisted AirComp studies therefore neglect this effect, which may lead to overly idealized performance evaluations \cite{Lyu2025PASSAirComp}.

\begin{figure}[!t]
\centering
    \subfigure[Segmented waveguide.]
    {
        \includegraphics[width=0.35\textwidth]{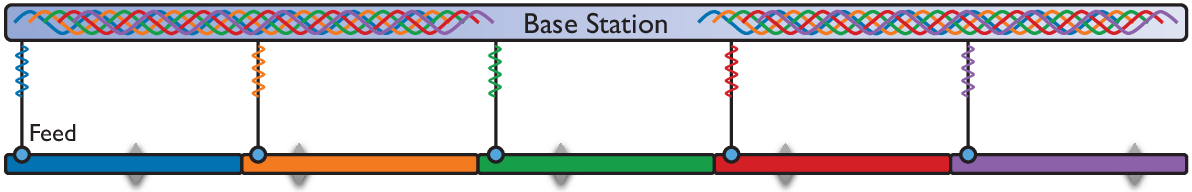}
	   \label{Figure: PAS_System_Model2}
    }
    \subfigure[System setup.]
    {
        \includegraphics[width=0.35\textwidth]{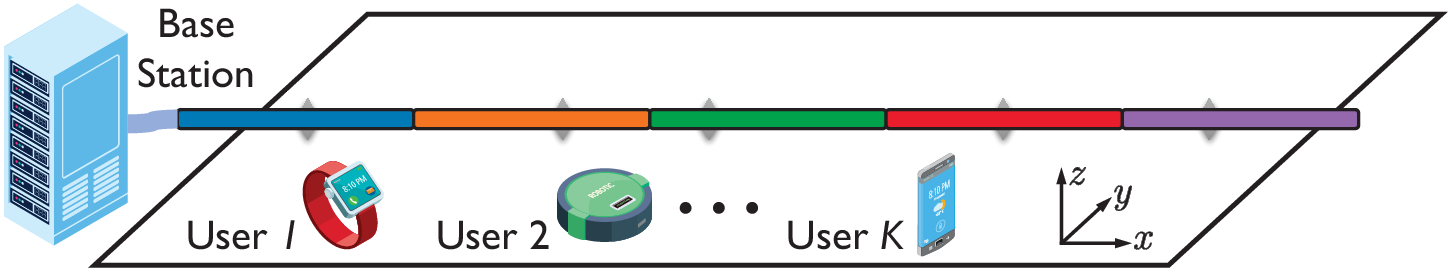}
	   \label{Figure: PAS_System_Model1}
    }
\caption{Illustration of the SWAN-based AirComp.}
\label{Figure: PAS_System_Model}
\vspace{-15pt}
\end{figure}

\begin{figure}[!t]
\centering
    \subfigure[SS.]
    {
        \includegraphics[width=0.145\textwidth]{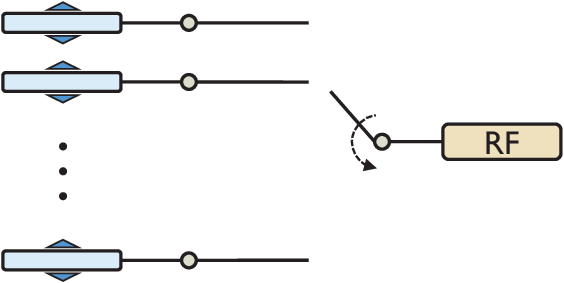}
	   \label{Figure: PAN_Protocol1}
    }
    \subfigure[SA (Type I).]
    {
        \includegraphics[width=0.145\textwidth]{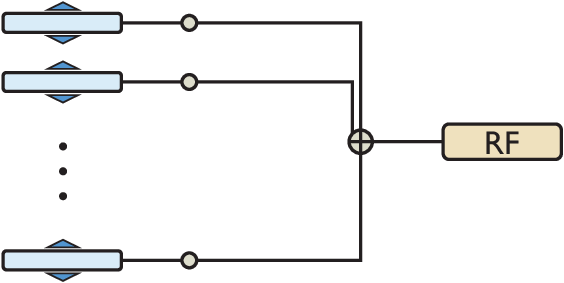}
	   \label{Figure: PAN_Protocol2}
    }
    \subfigure[SA (Type II).]
    {
        \includegraphics[width=0.145\textwidth]{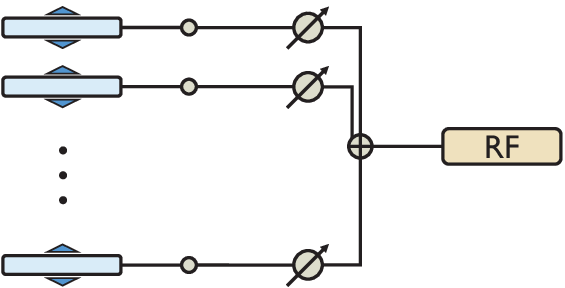}
	   \label{Figure: PAN_Protocol5}
    }
\caption{Illustration of the SWAN architectures.}
\label{Figure: PAS_Architecture}
\vspace{-15pt}
\end{figure}

To suppress IAR, the segmented waveguide-enabled pinching-antenna system (SWAN) was proposed in \cite{Ouyang2026SWAN}. SWAN partitions a long waveguide into multiple independent segments and activates at most one PA on each segment; see {\figurename} {\ref{Figure: PAS_System_Model2}}. This architecture suppresses IAR while preserving the spatial flexibility of PASS. Motivated by these advantages, this article studies SWAN-assisted AirComp and optimizes the PA deployment under three proposed transmission architectures. The first architecture is segment selection (SS), where only one segment is activated and connected to the radio-frequency (RF) chain through a switching network. The second architecture is segment aggregation (SA) without phase shifters (Type-I SA), where all segments are simultaneously activated and directly combined at the RF front end. The third architecture is SA with phase shifters (Type-II SA), where segment-wise phase control is further introduced before RF combining. These architectures are illustrated in {\figurename} {\ref{Figure: PAS_Architecture}}. For all architectures, we minimize the computation mean-squared error (MSE) through the joint design of the receive scaling factor, the PA locations, and the segment phase shifts. For SS, we develop a two-stage search algorithm to approach the optimal PA deployment. For SA, we propose low-complexity element-wise optimization algorithms to obtain high-quality solutions. Numerical results demonstrate that SWAN significantly improves AirComp accuracy compared with conventional PASS architectures. Moreover, segment-wise phase control further enhances channel alignment and aggregation performance.

\section{System Model}
Consider uplink transmission in a SWAN-assisted AirComp system, where $K$ single-antenna users simultaneously transmit their local data to a base station (BS) for function aggregation, as illustrated in {\figurename}~\ref{Figure: PAS_System_Model2}. The users are distributed within a rectangular service region with dimensions $D_x$ and $D_y$ along the $x$- and $y$-axes, respectively. The location of user $k$ is denoted by ${\mathbf{u}}_k=[u_k^x,u_k^y,0]^{\mathsf T}$ for $k\in{\mathcal K}\triangleq\{1,\ldots,K\}$. As shown in {\figurename}~\ref{Figure: PAS_System_Model1}, the segmented waveguide is deployed along the $x$-axis to provide horizontal coverage, while the PAs placed on the waveguide are used to receive uplink signals.

The segmented waveguide consists of $M$ independent dielectric waveguide segments, each with length $L$. Let ${\bm\psi}_0^m\triangleq[\psi_0^m,0,d]^{\mathsf T}$ denote the feed-point location of the $m$th segment for $m\in{\mathcal M}\triangleq\{1,\ldots,M\}$, where $\psi_0^1<\psi_0^2<\cdots<\psi_0^M$, and $d$ denotes the deployment height of the waveguide. For simplicity, the feed point is placed at the left end of each segment. To suppress uplink IAR, at most one PA is activated on each segment. The location of the PA on the $m$th segment is denoted by ${\bm\psi}_m\triangleq[\psi_m,0,d]^{\mathsf T}$ and satisfies
\begin{align}\label{PA_Location_Constraint}
\psi_{0}^{m}\leq \psi_{m}\leq \psi_{0}^{m}+L,\lvert\psi_{m}-\psi_{m'}\rvert\geq \Delta,\forall m\ne m',
\end{align}
where $\Delta>0$ denotes the minimum inter-antenna spacing required to suppress mutual coupling.
\subsection{Channel Model}
PASS is mainly envisioned for high-frequency communications \cite{suzuki2022pinching}, where LoS propagation dominates \cite{ouyang2024primer}. A free-space LoS channel model is therefore adopted. Under this model, the spatial channel coefficient between user $k$ and the PA on the $m$th segment is given by \cite{ouyang2024primer}
\begin{align}
h_{\rm o}({\mathbf u}_k,{\bm\psi}_m)
\triangleq
\frac{\eta^{\frac{1}{2}}
{\rm e}^{-{\rm j}k_0\|{\mathbf u}_k-{\bm\psi}_m\|}}
{\|{\mathbf u}_k-{\bm\psi}_m\|},
\end{align}
where $\eta\triangleq\frac{c^2}{16\pi^2f_{\rm c}^2}$, $c$ denotes the speed of light, $f_{\rm c}$ denotes the carrier frequency, $\lambda$ is the free-space wavelength, and $k_0=\frac{2\pi}{\lambda}$ is the wavenumber.

The in-waveguide propagation coefficient between the feed point and the PA on the $m$th segment follows \cite{pozar2021microwave}
\begin{align}\label{In_Waveguide_Channel_Model}
h_{\rm i}({\bm\psi}_m,{\bm\psi}_0^m)
&\triangleq
10^{-\frac{\kappa}{20}\|{\bm\psi}_m-{\bm\psi}_0^m\|}
{\rm e}^{-{\rm j}\frac{2\pi\|{\bm\psi}_m-{\bm\psi}_0^m\|}{\lambda_{\rm g}}},
\end{align}
where $\lambda_{\rm g}=\frac{\lambda}{n_{\rm eff}}$ denotes the guided wavelength, $n_{\rm eff}$ denotes the effective refractive index of the dielectric waveguide, and $\kappa$ denotes the average attenuation factor in dB/m. The special case $\kappa=0$ corresponds to a lossless dielectric waveguide. Prior studies showed that in-waveguide attenuation has negligible impact on SWAN performance \cite{Ouyang2026SWAN}. Based on this observation, we set $\kappa=0$ in the proposed design. The impact of nonzero attenuation is further evaluated in Section~\ref{Section: Numerical Results}.
\subsection{Transmission Architectures}
The performance of SWAN-assisted AirComp depends on the adopted transmission architecture, which determines how the segment feed points connect to the RF chain. In this work, three representative architectures are proposed.
\subsubsection{Segment Selection}
Under SS, shown in {\figurename}~\ref{Figure: PAN_Protocol1}, only one segment is connected to the RF chain through a switching network. Let $m\in{\mathcal M}$ denote the index of the selected segment. The effective uplink channel of user $k$ is then given by
\begin{subequations}
\begin{align}
h_k&=
h_{\rm i}({\bm\psi}_{ m},{\bm\psi}_0^{ m})
h_{\rm o}({\mathbf u}_k,{\bm\psi}_{ m})\\
&=\frac{\eta^{\frac{1}{2}}{\rm{e}}^{-{\rm{j}}k_0(\sqrt{(u_k^x-{\psi}_{m})^2+d_k}+n_{\rm{eff}}({\psi}_{m}-{\psi}_{0}^{m})) }}{\sqrt{(u_k^x-{\psi}_{m})^2+d_k}},
\end{align}
\end{subequations}
where $d_k\triangleq d^2+(u_k^y)^2$.
\subsubsection{Type-I Segment Aggregation}
To exploit the spatial array gain provided by multiple segments, we further consider Type-I SA, shown in {\figurename}~\ref{Figure: PAN_Protocol2}, where all segment feed points are simultaneously connected to a single RF chain through a signal combiner. The effective uplink channel of user $k$ is given by
\begin{subequations}\label{Channel_Type_I_SA}
\begin{align}
h_k&=
\frac{1}{\sqrt{M}}
\sum_{m=1}^{M}
h_{\rm i}({\bm\psi}_m,{\bm\psi}_0^m)
h_{\rm o}({\mathbf u}_k,{\bm\psi}_m)\\
&=\frac{\eta^{\frac{1}{2}}}{\sqrt{M}}
\sum_{m=1}^{M}\frac{{\rm{e}}^{-{\rm{j}}k_0(\sqrt{(u_k^x-{\psi}_{m})^2+d_k}+n_{\rm{eff}}({\psi}_{m}-{\psi}_{0}^{m})) }}{\sqrt{(u_k^x-{\psi}_{m})^2+d_k}},
\end{align}
\end{subequations}
where the normalization factor $\frac{1}{\sqrt{M}}$ accounts for the cumulative receiver noise introduced by the $M$ segments \cite{Ouyang2026SWAN}.
\subsubsection{Type-II Segment Aggregation}
To further improve channel alignment, we also consider Type-II SA, shown in {\figurename}~\ref{Figure: PAN_Protocol5}, where each segment is additionally equipped with a phase shifter before RF combining. Compared with Type-I SA, this architecture provides additional phase-control flexibility. The effective uplink channel of user $k$ is therefore given by
\begin{subequations}\label{Channel_Type_II_SA}
\begin{align}
h_k&=
\frac{1}{\sqrt{M}}
\sum_{m=1}^{M}
{\rm e}^{{\rm j}\theta_m}
h_{\rm i}({\bm\psi}_m,{\bm\psi}_0^m)
h_{\rm o}({\mathbf u}_k,{\bm\psi}_m)\\
&=\sum_{m=1}^{M}\frac{\eta^{\frac{1}{2}}{\rm{e}}^{{\rm j}\theta_m-{\rm{j}}k_0(\sqrt{(u_k^x-{\psi}_{m})^2+d_k}+n_{\rm{eff}}({\psi}_{m}-{\psi}_{0}^{m})) }}{\sqrt{M}\sqrt{(u_k^x-{\psi}_{m})^2+d_k}},
\end{align}
\end{subequations}
where $\theta_m\in[0,2\pi)$ denotes the phase shift applied to the $m$th segment.

\subsection{Computation Model}
Regardless of the adopted transmission architecture, the received signal at the BS can be expressed as follows:
\begin{align}
y=
\sum_{k=1}^{K}
\sqrt{P}h_kx_k+n,
\end{align}
where $x_k$ denotes the normalized local data symbol of user $k$ with $\mathbbmss{E}[x_k]=0$ and $\mathbbmss{E}[|x_k|^2]=1$, $P$ denotes the per-user transmit power, and $n\sim{\mathcal{CN}}(0,\sigma^2)$ denotes additive white Gaussian noise with covariance $\sigma^2$. Following the classical AirComp model, the BS aims to recover the summation function $x=\sum_{k=1}^{K}x_k$ through the waveform superposition property of the wireless MAC \cite{Wang2024AirComp}. It is assumed that the user data symbols are mutually independent.

The BS applies a receive scaling factor $r\in{\mathbbmss C}$ to estimate the summation function as $\hat{x}=ry=r\sum_{k=1}^{K}\sqrt{P}h_kx_k+rn$. The computation accuracy is evaluated through the mean-squared error (MSE) defined as follows:
\begin{align}\label{MSE_Initial}
{\mathsf{MSE}}
=
\mathbbmss{E}\!\left[|\hat{x}-x|^2\right]
=
\sum_{k=1}^{K}
|r\sqrt{P}h_k-1|^2
+
|r|^2\sigma^2.
\end{align}
Since \eqref{MSE_Initial} is convex with respect to $r$, the optimal receive scaling factor can be obtained by setting $\frac{\partial\mathsf{MSE}}{\partial r}=0$, which yields $r^\star=
\frac{\sum_{k=1}^{K}\sqrt{P}h_k^*}{\sum_{k=1}^{K}P|h_k|^2+\sigma^2}$. Substituting $r=r^\star$ into \eqref{MSE_Initial} gives the minimum MSE as follows:
\begin{align}
\mathsf{MSE}
=
K-
\frac{
P\lvert\sum_{k=1}^{K}h_k\rvert^2
}
{
\sum_{k=1}^{K}P|h_k|^2+\sigma^2
}.
\end{align}

Since the effective channel $h_k$ depends on the PA locations and the phase shifts, the next section develops low-complexity optimization algorithms to jointly design the PA deployment and segment phase shifts for SWAN-assisted AirComp under the three considered transmission architectures.

\section{Proposed Solution}

The MSE minimization problem is non-convex since the effective channels contain highly oscillatory distance-dependent phase terms. Exhaustive search over all PA locations and segment phase shifts therefore incurs prohibitive complexity. In this section, low-complexity optimization algorithms are developed for the three SWAN transmission architectures.

\subsection{Segment Selection}
Under SS, only one segment is connected to the RF chain. The resulting MSE can be expressed as follows:
\begin{align}
\mathsf{MSE}(\psi_m)
=
K-
\frac{
P\eta\left|
\sum_{k=1}^{K}
\frac{
{\rm e}^{-{\rm j}k_0\sqrt{(u_k^x-\psi_m)^2+d_k}}
}
{
\sqrt{(u_k^x-\psi_m)^2+d_k}
}
\right|^2
}
{
\sum_{k=1}^{K}
P\frac{\eta}{(u_k^x-\psi_m)^2+d_k}
+\sigma^2
}.
\end{align}

The design problem reduces to selecting one segment and optimizing one PA location within that segment. Exhaustive search requires $MQ$ MSE evaluations, where $Q$ denotes the number of candidate PA locations per segment. To reduce complexity, a two-stage search strategy is adopted. In the first stage, the midpoint of each segment is evaluated. Specifically, the PA on segment $m$ is initialized as $\psi_m^{\rm mid}=\psi_0^m+\frac{L}{2}$. The selected segment is given by
\begin{align}
m^\star
=
\argmin\nolimits_{m\in{\mathcal M}}
\mathsf{MSE}(\psi_m^{\rm mid}).
\end{align}
In the second stage, a refined search is performed only within the selected segment. Define the candidate set ${\mathcal Q}_{m^\star}
\triangleq\{\psi_0^{m^\star}+\frac{qL}{Q-1}|q=0,1,\ldots,Q-1\}$. The PA location is then updated as follows:
\begin{align}
\psi^\star
=
\argmin\nolimits_{x\in{\mathcal Q}_{m^\star}}
\mathsf{MSE}(x).
\end{align}
This two-stage procedure reduces the complexity from ${\mathcal O}(KMQ)$ to ${\mathcal O}(K(M+Q))$.
\subsection{Segment Aggregation}
\subsubsection{Type-I Segment Aggregation}
Under Type-I SA, all segments are connected to the RF chain without phase shifters. The effective channel is given in \eqref{Channel_Type_I_SA}. The optimization problem is formulated as follows:
\begin{align}
\min\nolimits_{\{\psi_m\}_{m=1}^{M}}
~
\mathsf{MSE}
\quad
{\rm s.t.}
\quad
\eqref{PA_Location_Constraint}.
\end{align}

This problem is difficult to solve globally due to the non-convex propagation phases. However, when all PA locations except $\psi_m$ are fixed, the resulting subproblem becomes one-dimensional. This observation motivates an element-wise alternating-optimization (AO) framework, where each PA location is updated sequentially. The subproblem for optimizing $\psi_m$ can be formulated as follows:
\begin{subequations}
\begin{align}
\max_{\psi_m}~&f_m(\psi_m)\triangleq\frac{
\frac{P\eta}{M}\lvert\sum_{k=1}^{K}[\hat{g}_k(\psi_m)+\sum_{m'\ne m}\hat{g}_k^m]\rvert^2
}
{
\sum_{k=1}^{K}\frac{P\eta}{M}|\hat{g}_k(\psi_m)+\sum_{m'\ne m}\hat{g}_k^m|^2+\sigma^2
}\nonumber\\{\rm{s.t.}}~&{\psi}_{m}\in[\psi_{0}^{m},\psi_{0}^{m}+L],\lvert{\psi}_{m}-{\psi}_{m'}\rvert\geq\Delta,m\ne m'\nonumber,
\end{align}
\end{subequations}
where $\hat{g}_k(\psi_m)\triangleq\frac{{\rm{e}}^{-{\rm{j}}k_0(({(u_k^x-{\psi}_{m})^2+d_k})^{1/2}+n_{\rm{eff}}({\psi}_{m}-{\psi}_{0}^{m})) }}{({(u_k^x-{\psi}_{m})^2+d_k})^{1/2}}$ and $\hat{g}_k^m\triangleq\sum_{m'\ne m}\frac{{\rm{e}}^{-{\rm{j}}k_0(({(u_k^x-{\psi}_{m'})^2+d_k})^{1/2}+n_{\rm{eff}}({\psi}_{m'}-{\psi}_{0}^{m'})) }}{({(u_k^x-{\psi}_{m'})^2+d_k})^{1/2}}$. Since the subproblem contains only one scalar variable over a bounded interval, it can be efficiently solved through one-dimensional search. The feasible interval is discretized into ${\mathcal Q}_m=\{\psi_0^m+\frac{qL}{Q-1}|q=0,\ldots,Q-1\}$. Due to the minimum-spacing constraint, infeasible points are excluded through
\begin{align}
\hat{\mathcal{Q}}_m\triangleq \{x|x\in{\mathcal{Q}}_m,\lvert x - \psi_{m'}\rvert<\Delta,m'\ne m\}.
\end{align}
The PA location update is therefore given by
\begin{align}
\psi_m^\star
=
\argmax\nolimits_{\psi_m\in{\mathcal Q}_m\setminus\hat{\mathcal Q}_m}
f_m(\psi_m).
\end{align}
The above updates are repeated sequentially over all segments until convergence.
\subsubsection{Type-II Segment Aggregation}
Under Type-II SA, each segment is additionally equipped with a phase shifter. The effective channel is given in \eqref{Channel_Type_II_SA}. Both the PA locations and the phase shifts need to be optimized. Since these two variable blocks are strongly coupled, an AO framework is adopted. The PA locations are updated through the same one-dimensional search as in Type-I SA, while the phase shifts admit closed-form element-wise updates.

For fixed $\{\psi_m\}_{m=1}^{M}$ and $\{\theta_{m'}\}_{m'\neq m}$, define $a_k \triangleq
\frac{1}{\sqrt{M}}
h_{\rm i}({\bm\psi}_m,{\bm\psi}_0^m)
h_{\rm o}({\mathbf u}_k,{\bm\psi}_m)$ and $b_k \triangleq
\sum_{m'\ne m}
\frac{{\rm e}^{{\rm j}\theta_{m'}}}{\sqrt{M}}
h_{\rm i}({\bm\psi}_\ell,{\bm\psi}_0^{m'})
h_{\rm o}({\mathbf u}_k,{\bm\psi}_{m'})$. Then $h_k=b_k+{\rm e}^{{\rm j}\theta_m}a_k$. The phase-update subproblem becomes
\begin{align}\label{Phase_Objective_New}
\max\nolimits_{\theta_m\in[0,2\pi)}\phi(\theta_m)
=
\frac{
\left|
\sum_{k=1}^{K}
(b_k+{\rm e}^{{\rm j}\theta_m}a_k)
\right|^2
}
{
\sum_{k=1}^{K}
|b_k+{\rm e}^{{\rm j}\theta_m}a_k|^2
+\frac{\sigma^2}{P}
}.
\end{align}

Define
\begin{align}
A&=\sum_{k=1}^{K}a_k,
\quad
B=\sum_{k=1}^{K}b_k,
\quad
C=\sum_{k=1}^{K}a_kb_k^*,
\end{align}
and express
\begin{align}
AB^*=|\beta|{\rm e}^{{\rm j}p},
\quad
C=|\delta|{\rm e}^{{\rm j}q}.
\end{align}
Then \eqref{Phase_Objective_New} can be rewritten as follows:
\begin{align}
\max\nolimits_{\theta_m\in[0,2\pi)}\phi(\theta_m)
=
\frac{
\alpha+2|\beta|\cos(\theta_m+p)
}
{
\gamma+2|\delta|\cos(\theta_m+q)
},
\end{align}
where $\alpha=|A|^2+|B|^2$ and $\gamma=\sum_{k=1}^{K}(|a_k|^2+|b_k|^2)+\frac{\sigma^2}{P}$.

By setting $\frac{{\rm d}\phi(\theta_m)}{{\rm d}\theta_m}=0$, we obtain
\begin{align}
A_m\sin\theta_m+B_m\cos\theta_m=\xi_m,
\end{align}
where
\begin{align}
A_m&=
2\gamma|\beta|\cos p
-
2\alpha|\delta|\cos q,
\\
B_m&=
2\gamma|\beta|\sin p
-
2\alpha|\delta|\sin q,
\\
\xi_m&=
4|\beta||\delta|\sin(q-p).
\end{align}

Let $R_m=\sqrt{A_m^2+B_m^2}$ and $\varphi_m=\operatorname{atan2}(A_m,B_m)$. The stationary points are therefore given by
\begin{align}
\theta_m^{(1)}
&=
\varphi_m+
\arccos\left(\frac{\xi_m}{R_m}\right),
\\
\theta_m^{(2)}
&=
\varphi_m-
\arccos\left(\frac{\xi_m}{R_m}\right).
\end{align}
The optimal phase update is obtained as
\begin{align}
\theta_m^\star
=
\argmax\nolimits_{\theta\in\{\theta_m^{(1)},\theta_m^{(2)}\}}
\phi(\theta).
\end{align}
This closed-form update avoids exhaustive phase search and reduces the phase-update complexity to ${\mathcal O}(K)$ per segment. The overall AO procedure for Type-II SA is summarized in Algorithm~\ref{alg:ao_typeII}.

\begin{algorithm}[t]
\algsetup{linenosize=\tiny} \scriptsize
\caption{AO for Type-II SA}
\label{alg:ao_typeII}
\begin{algorithmic}[1]
\REQUIRE User locations $\{\mathbf u_k\}_{k=1}^{K}$, segment set ${\mathcal M}$, position grid size $Q$
\ENSURE PA locations $\{\psi_m\}$ and phase shifts $\{\theta_m\}$
\STATE Initialize $\psi_m=\psi_0^m+L/2$ and $\theta_m=0$, $\forall m$
\REPEAT
\FOR{$m=1,\ldots,M$}
\STATE Update $\theta_m$ by the closed-form element-wise phase solution
\ENDFOR
\FOR{$m=1,\ldots,M$}
\STATE Update $\psi_m$ by one-dimensional search over ${\mathcal Q}_m$
\ENDFOR
\UNTIL{convergence}
\RETURN $\{\psi_m\}$ and $\{\theta_m\}$
\end{algorithmic}
\end{algorithm}

\subsubsection{Convergence and Complexity}
The proposed algorithms are based on element-wise alternating optimization. For Type-I SA, each PA-location update globally optimizes the corresponding one-dimensional subproblem over the discretized feasible set while all remaining variables are fixed. For Type-II SA, each phase-shift update globally optimizes the corresponding element-wise phase subproblem in closed form, while each PA-location update again globally optimizes the corresponding one-dimensional search subproblem. Therefore, the MSE is non-increasing after every iteration. Moreover, since the MSE is lower bounded by zero, the proposed AO algorithms are guaranteed to converge to a stationary point of the discretized optimization problem.

For Type-I SA, each AO iteration requires updating $M$ PA locations, where each update evaluates $Q$ candidate grid points and each evaluation requires ${\mathcal O}(K)$ operations. The resulting per-iteration complexity is therefore ${\mathcal O}(KMQ)$. For Type-II SA, the closed-form phase updates require ${\mathcal O}(KM)$ operations per iteration, while the PA-location updates require ${\mathcal O}(KMQ)$ operations. Hence, the total per-iteration complexity is ${\mathcal O}(KM(Q+1))$, which is dominated by the one-dimensional PA-location search.
\section{Numerical Results}\label{Section: Numerical Results}

This section evaluates the proposed SWAN-assisted AirComp schemes. Unless otherwise specified, we set $f_c=28$ GHz, $n_{\rm eff}=1.4$, $\Delta=\lambda/2$, $d=3$ m, $P=10$ dBm, and $\sigma^2=-90$ dBm. The grid resolution is $Q=10^3$ with random initialization. The users are uniformly distributed in a rectangular region with $D_x=100$ m and $D_y=20$ m, and the waveguide is centered above the service region.

\begin{figure}[t]
    \centering
    \includegraphics[width=0.48\textwidth]{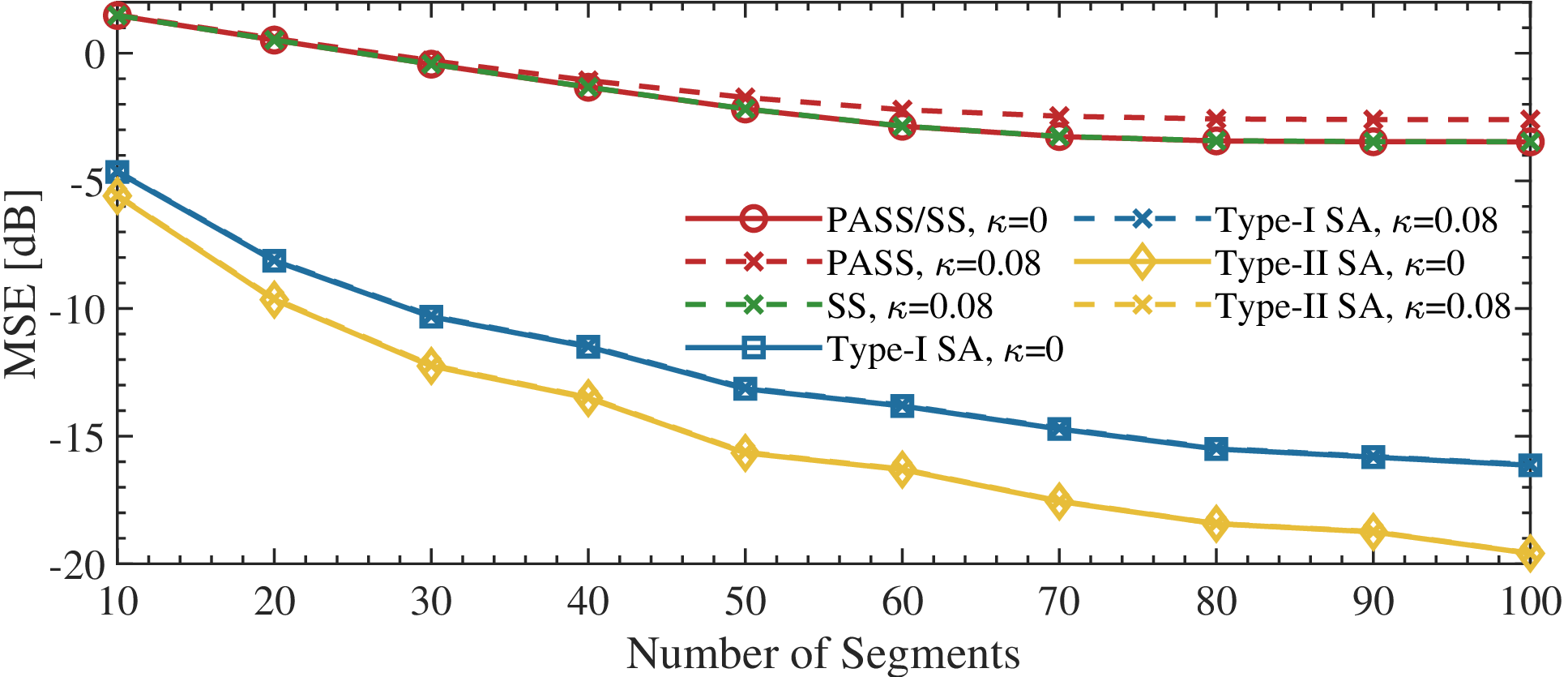}
    \caption{MSE versus the number of segments with $L=1$ m.}
    \label{Fig_MSE_vs_M_Fixed_L1}
\end{figure}

{\figurename} {\ref{Fig_MSE_vs_M_Fixed_L1}} shows the MSE versus the number of segments for fixed segment length $L=1$ m. The conventional PASS benchmark uses a single continuous waveguide and one PA, since a tractable multi-PA uplink model for conventional PASS is unavailable due to IAR. Both lossless waveguides with $\kappa=0$ and lossy waveguides with $\kappa=0.08$ dB/m are considered \cite{Ouyang2026SWAN}. When $\kappa=0$, SS achieves the same performance as the conventional single-PA PASS because both schemes use only one active PA. When $\kappa>0$, SS outperforms conventional PASS since SWAN confines the in-waveguide propagation loss within each short segment, while conventional PASS suffers from waveguide-level attenuation. For SA, Type-I SA achieves much lower MSE than SS due to multi-PA aggregation gain, while Type-II SA further improves the performance through segment-wise phase control. The results also show that moderate in-waveguide attenuation has only a limited impact on SWAN, which validates the robustness of the segmented architecture.

\begin{figure}[t]
    \centering
    \includegraphics[width=0.48\textwidth]{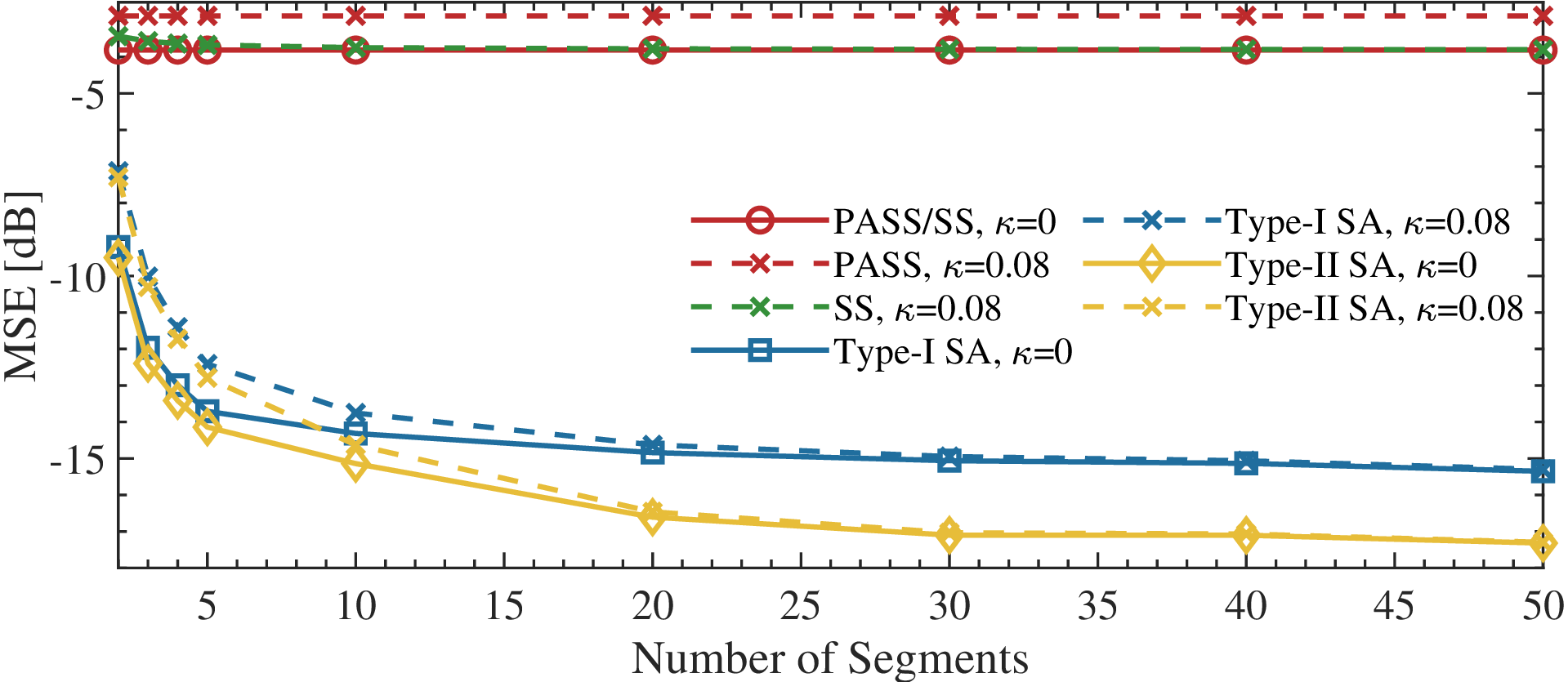}
    \caption{MSE versus the number of segments with fixed waveguide length $ML=D_x$.}
    \label{Fig_MSE_vs_M_Fixed_Dx}
\end{figure}

{\figurename} {\ref{Fig_MSE_vs_M_Fixed_Dx}} further studies the case with fixed total waveguide length $D_x$, where $L=D_x/M$. In this case, increasing $M$ shortens each segment. As a result, the lossy SWAN schemes improve as $M$ increases, since the average in-waveguide propagation distance is reduced. The performance gap between the lossy and lossless SWAN curves therefore becomes smaller for large $M$. The SA schemes also benefit from the increased number of active PAs. Type-II SA consistently provides the lowest MSE, which confirms the value of phase-controlled segment aggregation.

\begin{figure}[t]
    \centering
    \includegraphics[width=0.48\textwidth]{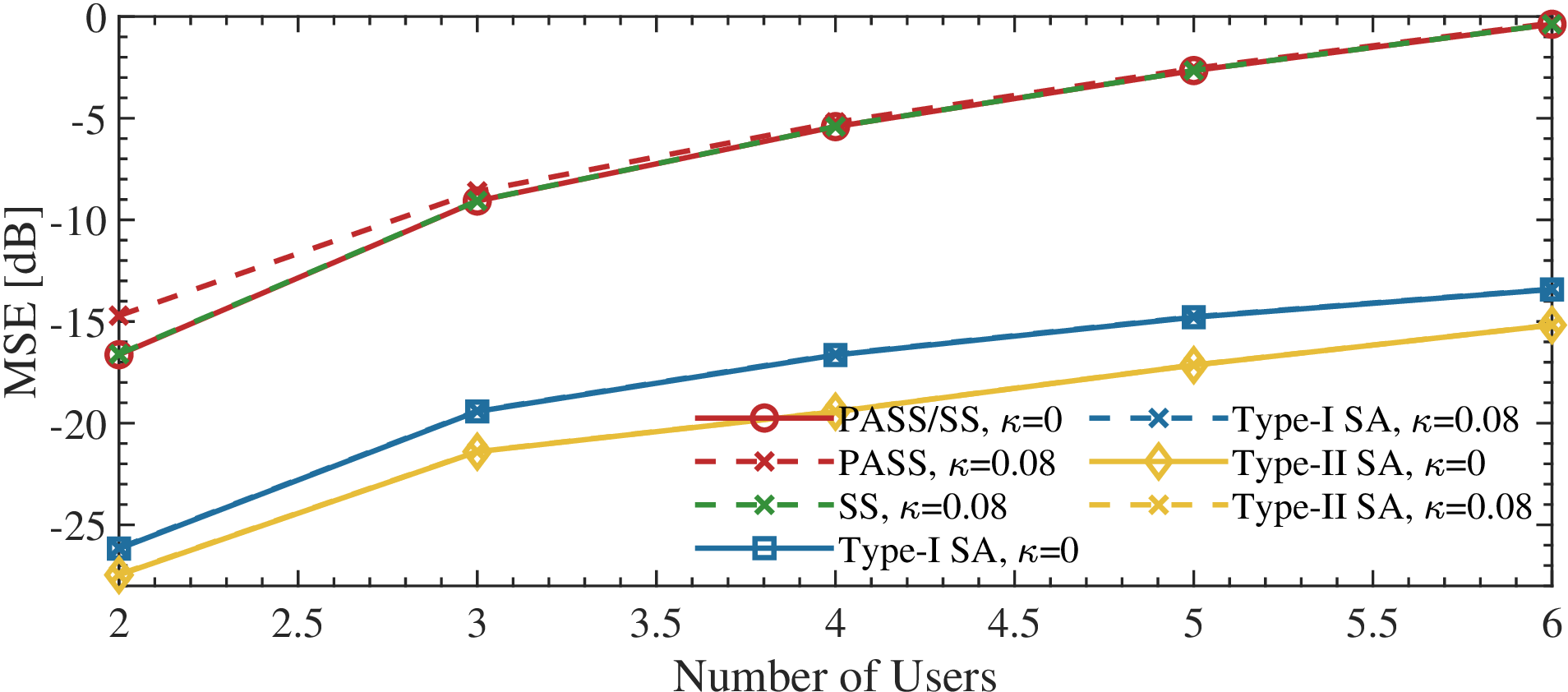}
    \caption{MSE versus the number of users with $D_x=50$ m.}
    \label{Fig_MSE_vs_K_Fixed_L}
\end{figure}

{\figurename} {\ref{Fig_MSE_vs_K_Fixed_L}} plots the MSE versus the number of users. The MSE increases with $K$ for all schemes, since channel alignment becomes more difficult when more distributed signals are superimposed. Nevertheless, the proposed SWAN schemes outperform conventional PASS, especially in the lossy case. This is because segmentation reduces in-waveguide attenuation and provides more flexible PA deployment. Among the considered schemes, Type-II SA achieves the best performance due to its additional phase-alignment capability.

\begin{figure}[t]
    \centering
    \includegraphics[width=0.48\textwidth]{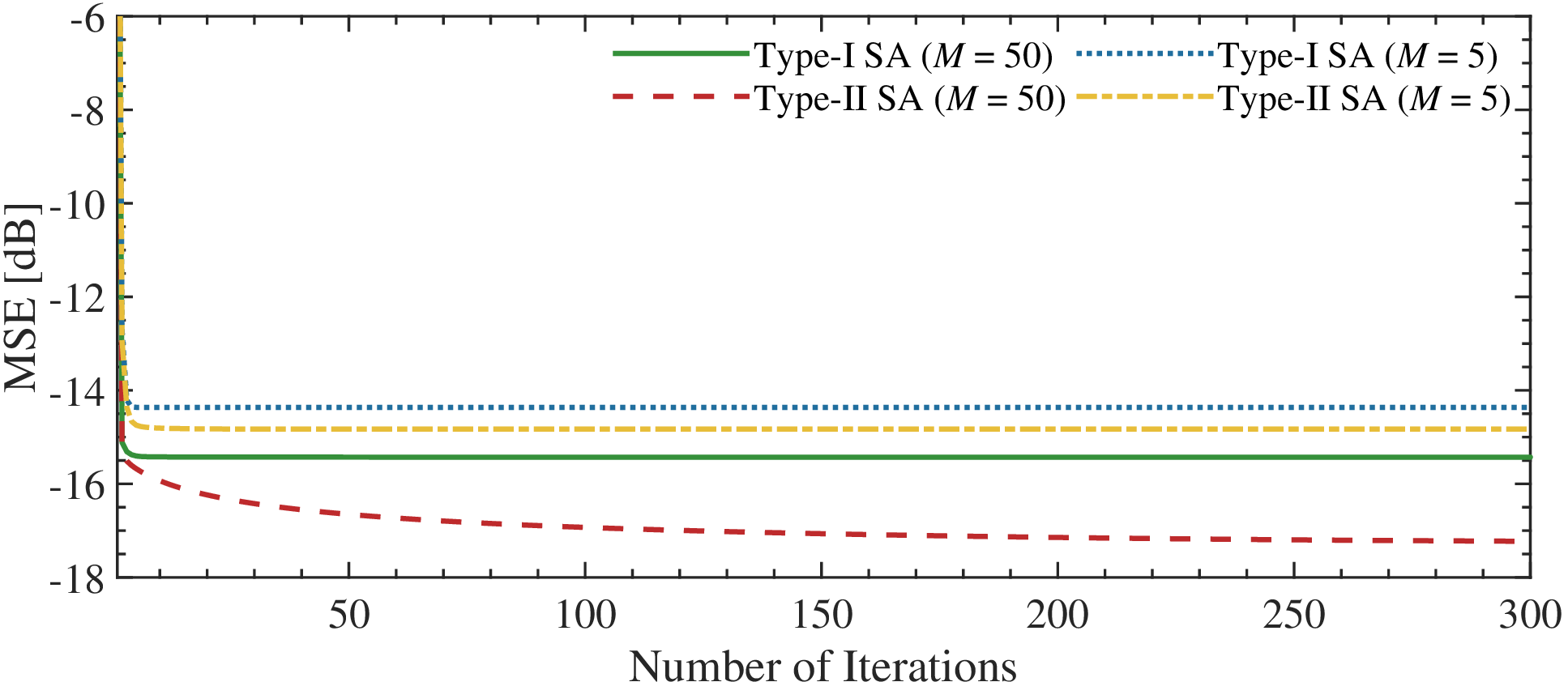}
    \caption{Convergence of the proposed algorithms. $D_x=50$ m.}
    \label{fig_conv}
\end{figure}

{\figurename} {\ref{fig_conv}} shows the convergence behavior of the proposed AO algorithms for SA. Both Type-I and Type-II SA converge within a moderate number of iterations. Type-II SA reaches a lower MSE than Type-I SA, which demonstrates the benefit of joint PA placement and segment-wise phase control.

\section{Conclusion}
This article investigated SWAN-assisted AirComp for edge intelligence systems. We considered SS and SA transmission architectures and optimized the receive scaling factor, PA locations, and segment phase shifts to minimize the computation MSE. We developed a two-stage search method for SS and low-complexity AO algorithms for SA. For Type-II SA, we further derived closed-form element-wise phase updates to avoid exhaustive phase search. Our numerical results showed that SWAN effectively suppresses the impact of in-waveguide propagation loss through segmentation and significantly outperforms conventional PASS in lossy environments. The results also demonstrated that SA achieves substantial aggregation gains over SS, while segment-wise phase control in Type-II SA further improves channel alignment and reduces the MSE. These findings suggest that SWAN can provide a flexible and efficient physical-layer architecture for future compute-while-communicate systems.

\bibliographystyle{IEEEtran}
\bibliography{mybib}
\end{document}